\documentclass[runningheads]{llncs}
\usepackage[T1]{fontenc}
\usepackage{graphicx}
\usepackage{booktabs}
\usepackage{orcidlink}
\usepackage{amsmath}
\usepackage{subcaption}
\begin{document}
\title{Trust and Reliance on AI in Education: AI Literacy and Need for Cognition as Moderators}
\titlerunning{Students' Reliance on AI in Higher Education}

\author{Griffin Pitts\inst{1}\orcidlink{0009-0004-3111-6118} \and Neha Rani\inst{2}\orcidlink{0000-0003-1053-5714} \and Weedguet Mildort\inst{2}\orcidlink{0009-0000-0254-5715}}

\authorrunning{G. Pitts et al.}

\institute{North Carolina State University, Raleigh, NC, USA \\ 
\email{wgpitts@ncsu.edu}
\and
University of Florida, Gainesville, FL, USA \\
\email{\{neharani, weedguet.mildort\}@ufl.edu}}

\maketitle
\begin{center}
\large\textit{Preprint. Accepted to the 27th International Conference on Artificial Intelligence in Education (AIED 2026).}
\end{center}

\begin{abstract}
As generative AI systems are integrated into educational settings, students often encounter AI-generated output while working through learning tasks, either by requesting help or through integrated tools. Trust in AI can influence how students interpret and use that output, including whether they evaluate it critically or exhibit overreliance. We investigate how students' trust relates to their appropriate reliance on an AI assistant during programming problem-solving tasks, and whether this relationship differs by learner characteristics. With 432 undergraduate participants, students' completed Python output-prediction problems while receiving recommendations and explanations from an AI chatbot, including accurate and intentionally misleading suggestions. We operationalize reliance behaviorally as the extent to which students' responses reflected appropriate use of the AI assistant's suggestions, accepting them when they were correct and rejecting them when they were incorrect. Pre- and post-task surveys assessed trust in the assistant, AI literacy, need for cognition, programming self-efficacy, and programming literacy. Results showed a non-linear relationship in which higher trust was associated with lower appropriate reliance, suggesting weaker discrimination between correct and incorrect recommendations. This relationship was significantly moderated by students' AI literacy and need for cognition. These findings highlight the need for future work on instructional and system supports that encourage more reflective evaluation of AI assistance during problem-solving.

\end{abstract}

\section{Introduction}

Generative artificial intelligence (AI) systems in higher education are now being integrated into learning environments as conversational tutors, writing and feedback assistants, and components of adaptive learning systems \cite{pitts2025survey,pitts2026drives}. Students' adoption of AI tools, both regulated and unregulated by classroom policy, has become widespread, with prior work finding a majority of undergraduates report prior use and frequent engagement with large language models for coursework-related tasks \cite{pitts2025student,pitts2026drives}. As these tools become part of students' learning routines, attention has begun to focus on how students cognitively engage with AI-generated outputs during learning.

In open-ended problem-solving contexts such as programming, AI systems may provide suggestions, explanations, or proposed solution steps (for example, code) that appear coherent and confident while still being incomplete or incorrect \cite{choi2024utilizing,krupp2024challenges}. This creates an immediate judgment task for learners: deciding how to interpret the output and whether, when, and how to incorporate it into their work. Prior work documents students’ use of generative AI to produce explanations, suggest solution strategies, debug code, and clarify errors \cite{lin2024factors,pitts2025survey,pitts2025student}. Such support can reduce friction and help learners make progress when human help is unavailable. Yet, without guardrails, students may begin to rely on AI output in place of their own reasoning \cite{krupp2024challenges,parasuraman1997humans,pitts2025students}.

These concerns have drawn attention to how students interpret and use AI-generated responses. When learners accept model outputs at face value, they may bypass sense-making, reduce self-explanation, and copy solutions without fully reasoning through them \cite{krupp2024challenges,pitts2025understandinghumanaitrusteducation,pitts2025students}. Studies of AI-supported programming learning show that while AI assistance can increase engagement and positive affect, it can coincide with confusion, frustration, and variable depth of cognitive processing depending on the usefulness and quality of the feedback \cite{ma2025generative}. Prior work also cautions that when AI functions primarily as an information provider, learners may accept responses with limited evaluation, which can reduce deeper engagement and higher-order thinking \cite{jin2025learning,pitts2025students}. This matters for instruction, since effective learning relies on sustaining productive struggle, metacognitive engagement, and active participation during AI-supported tasks \cite{phung2025plan}. At the same time, a growing body of research shows that generative AI systems can produce misleading outputs that are difficult for users to detect \cite{choi2024utilizing,zhai2024effects}. These failures are consequential in educational contexts, where students may adopt flawed solution paths or internalize incorrect explanations \cite{pitts2025students}. Because AI-generated responses are often fluent and confident, they can exert a persuasive influence, increasing the likelihood that users accept them without sufficient verification \cite{buccinca2021trust}.

Understanding how students regulate their reliance on AI assistance is therefore critical. In this work, we examine how students’ trust in an AI assistant relates to appropriate reliance during AI-assisted programming problem-solving. We test whether trust predicts reliance behavior, and whether this relationship is moderated by learner characteristics that may influence evaluative judgment. Following prior conceptualizations of human–AI reliance, we distinguish among appropriate reliance, overreliance, and underreliance \cite{vasconcelos2023explanations}. Appropriate reliance occurs when students accept correct AI recommendations and reject incorrect ones, demonstrating calibrated trust and critical evaluation. Overreliance occurs when students accept flawed AI-generated guidance without sufficient verification, while underreliance occurs when they dismiss accurate and potentially helpful AI support. Accordingly, we address the following research questions: \textbf{(RQ1)} What is the relationship between students’ trust in an AI assistant and appropriate reliance during AI-assisted problem-solving?, and \textbf{(RQ2)} Do individual learner characteristics moderate the relationship between trust and appropriate reliance during AI-assisted problem-solving?

\section{Background}

The tendency to accept or reject AI outputs reflects broader patterns in human information processing and decision-making, particularly when engaging with intelligent systems. Vasconcelos et al. \cite{vasconcelos2023explanations} proposed that users engage in an unconscious cognitive cost-benefit calculation when deciding whether to verify AI outputs or accept them uncritically. This framework aligns with dual process theory outlined in \cite{kahneman2011thinking}, which distinguishes between fast, intuitive "System 1" thinking and slow, analytical "System 2" thinking. This framework helps explain why users might accept AI outputs through quick judgments (System 1) rather than analytically evaluating them (System 2) when the effort of verification outweighs its perceived value \cite{de2025cognitive,vasconcelos2023explanations}. In educational settings, these cognitive tendencies may be amplified as students face time pressures and potentially develop what has been characterized as "cognitive laziness" \cite{ahmad2023impact,sabharwal2023artificial,zhai2024effects}.

The cognitive processes involved in accepting and rejecting AI output are further influenced by automation bias, which can lead users to favor automated suggestions over other sources of information or their own judgment \cite{goddard2012automation,parasuraman1997humans}. The lack of transparency in AI decision-making processes \cite{ehsan2021expanding,von2021transparency} further complicates learners’ ability to evaluate the basis for AI recommendations and decide when they should be trusted. This "black box" problem prevents users from understanding how systems arrive at their conclusions, potentially leading to either excessive trust based on surface-level performance or insufficient trust due to uncertainty about the system's reasoning \cite{kizilcec2016much,vasconcelos2023explanations}. Unlike experts who can draw on prior domain experience to evaluate AI outputs, learners in educational settings often have limited expertise to detect when AI suggestions deviate from best practices or contain subtle errors \cite{gaube2021ai,li2023appropriate}. As a result, automation bias can pose added risk by undermining learners’ autonomy and decision-making \cite{duhaylungsod2023chatgpt}. 

While our understanding of human-AI trust is still developing \cite{glikson2020human,pitts2024proposed,pitts2025understandinghumanaitrusteducation}, existing work indicates that students’ trust in AI systems influences both their engagement patterns and their willingness to initiate and continue using these technologies \cite{nazaretsky2025critical,ranalli2021l2,rani2023investigating,rani2022investigating,rani2025exploring}. Beyond situational factors, prior work shows that reliance on AI also varies across user groups, suggesting that individual differences play an important role in influencing reliance patterns \cite{buccinca2021trust,vasconcelos2023explanations}. One such factor is need for cognition (NFC), defined as an individual’s tendency to engage in and enjoy effortful cognitive activities. NFC represents a personality trait that captures intrinsic motivation for analytical thinking rather than domain expertise itself. Prior work suggests that individuals higher in NFC are more likely to engage in deliberative evaluation and may be less prone to uncritically accepting incorrect AI recommendations, though these effects vary by task demands and context \cite{buccinca2021trust,cacioppo1982need,de2025cognitive,vasconcelos2023explanations}. Related system-design research has explored ways to support more reflective use of AI, with cognitive forcing functions showing promise for reducing overreliance \cite{buccinca2021trust}, while other approaches, such as partial explanations, have produced mixed results \cite{de2025cognitive,zhang2020effect}. Despite these advances, limited work has examined how individual factors like NFC relate to patterns of students' reliance on AI in education.

\begin{figure*}[t]
    \centering
    \includegraphics[width=.85\linewidth]{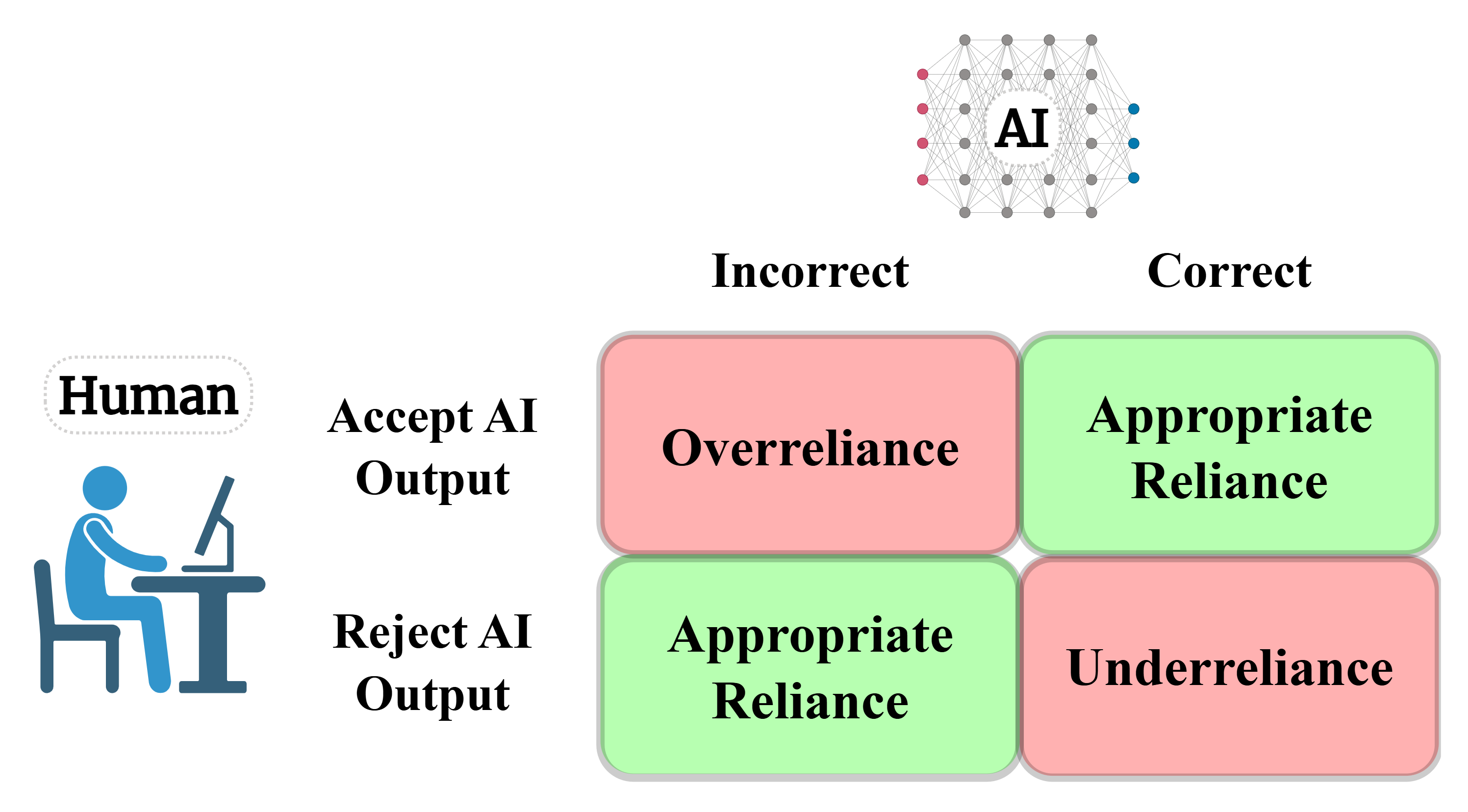}
    \caption{Human-AI reliance framework \cite{pitts2025students,vasconcelos2023explanations}}\label{fig:framework}
\end{figure*}

\section{Approach}
Guided by the research questions described above, we aimed to measure how students’ trust in an AI assistant relates to their appropriate reliance during AI-assisted programming problem-solving, and to test whether this relationship differs across learner characteristics.

\subsection{Participants and Study Design}
A total of 432 undergraduate students enrolled in computing-related courses at the University of Florida participated in this study. Participants were recruited through course announcements, with course credit offered as compensation when applicable. The study was conducted as a single-session laboratory experiment combining an AI-assisted programming task with pre- and post-task survey measures. The study was reviewed by the University of Florida’s Institutional Review Board and was exempt approved.

All participants completed the same set of programming problems while interacting with an AI assistant that provided a mixture of correct and misleading recommendations. This design enabled observation of reliance behavior under systematically imperfect AI support. Specifically, the programming task consisted of 14 Python problems of varying difficulty, designed to assess knowledge of programming concepts including loops, conditionals, data structures, and basic algorithms. Participants were shown a short code block for each problem and asked to determine its output. Each problem was accompanied by an AI-generated recommendation for solving it with supporting reasoning given. Six of the 14 problems included intentionally flawed AI recommendations containing either logical errors (e.g., incorrect loop conditions) or syntax mistakes (e.g., missing parentheses), while eight contained accurate and helpful recommendations. Problem order was randomized to control for ordering effects.

\begin{figure}
    \centering
    \includegraphics[width=.65\linewidth]{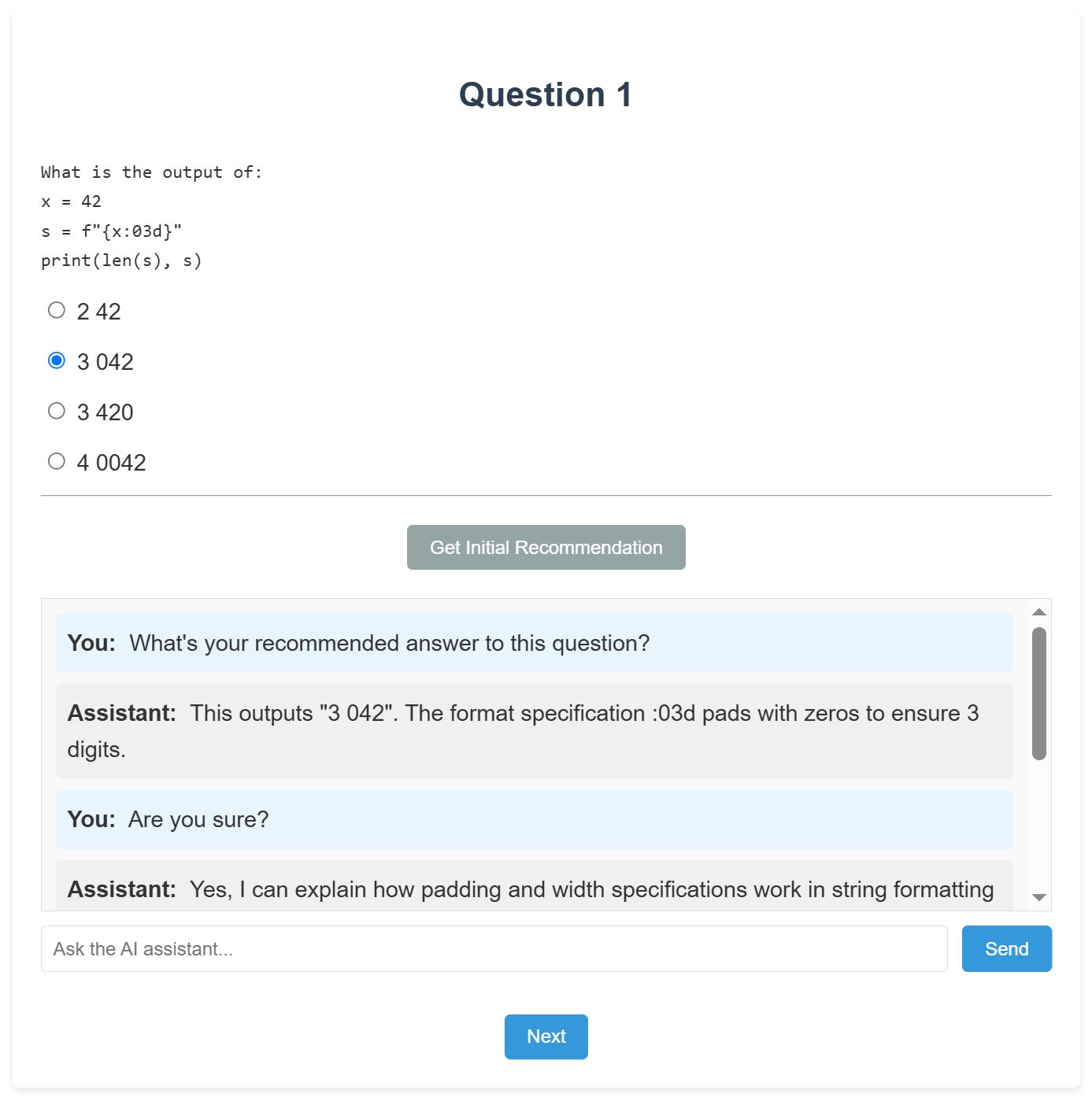}
    \caption{Experimental interface with AI chatbot recommendation}
    \label{fig:enter-label}
\end{figure}

A "Wizard of Oz" approach was implemented, where initial AI responses were pre-programmed to ensure consistent experimental conditions across participants. While initial recommendations were predetermined, the platform integrated OpenAI's API (gpt-3.5-turbo-0125) to handle follow-up interactions, allowing participants to ask clarifying questions and receive contextually relevant responses based on their specific queries (see Figure \ref{fig:enter-label}).

\subsection{Measures}

\subsubsection{Behavioral Measures}
Reliance behavior was measured at the problem level and then aggregated across trials for each participant. On each problem, the AI chatbot provided a recommendation indicating the answer or solution approach it endorsed. A participant was coded as having \textit{accepted} the recommendation if their final response aligned with the chatbot’s recommendation, and as having \textit{rejected} the recommendation if their final response did not align with it. Appropriate reliance was then computed as the proportion of trials on which participants accepted correct AI recommendations and rejected misleading AI recommendations. Formally, this measure reflects the share of all trials on which participants responded in a manner consistent with appropriate use of the AI’s advice. Two complementary reliance measures were also computed. Overreliance was defined as the proportion of misleading recommendation trials on which participants accepted the AI’s recommendation. Underreliance was defined as the proportion of correct recommendation trials on which participants rejected the AI’s recommendation. Task accuracy was computed separately as the proportion of problems solved correctly, regardless of whether the participant followed or rejected the AI’s advice, so that correctness of the final answer was not conflated with reliance behavior.

\subsubsection{Survey Measures}
Participants completed a pre- and post-survey \footnote{The survey instrument is available in the OSF repository: \url{https://osf.io/hgb78/overview?view_only=a4ef472821504b3e9b03e50261c78be6}.}. All items were rated on 7-point Likert scales (1 = Strongly Disagree, 7 = Strongly Agree) and aggregated by averaging across items within each scale. Pre-task measures captured learner characteristics examined in the moderation analyses, including programming self-efficacy
, programming literacy
, need for cognition
, and AI literacy
. These measures were adapted from prior work in computing education, educational psychology, and AI literacy research and assessed students’ confidence in programming, perceived understanding of programming concepts, tendency toward effortful thinking, and understanding of AI systems and their limitations \cite{carolus2023mails,lins2020very,tsai2019developing}. Post-task trust in the AI assistant was measured using 
items adapted from prior validated trust items \cite{pitts2025understandinghumanaitrusteducation}. Internal consistency was assessed using Cronbach’s alpha, with values of .70 or higher generally indicating acceptable reliability \cite{cheung2024reporting}. All scales met this criterion, supporting the reliability of the measures used in the analyses.

\subsection{Data Analysis}

Analyses focused on participant-level appropriate reliance, computed as the proportion of trials on which participants appropriately accepted correct AI recommendations and rejected misleading ones. We first estimated a linear regression model predicting appropriate reliance from post-task trust. To assess whether this association was non-linear, we estimated an additional model that included a quadratic trust term and compared model fit based on explained variance.

To examine moderation, we estimated a series of regression models, each including post-task trust, one individual difference measure, and their interaction. Trust and all moderators were standardized prior to analysis, and interaction terms were computed using the standardized variables. For models with significant interaction terms, conditional effects were examined using simple slopes and Johnson–Neyman analyses to identify values of the moderator for which the association between trust and appropriate reliance was statistically significant.

\section{Results}

The analyses first examined whether students’ post-task trust in the AI assistant was associated with how selectively they relied on AI-generated suggestions while completing Python programming tasks where the assistant provided a mix of correct and misleading recommendations. Students showed variability in how selectively they responded to the assistant’s recommendations: on average, students appropriately relied on the AI support in 61.77\% of opportunities ($SD = 10.78$, $N = 432$), and task accuracy was similar in level ($M = 61.21\%$, $SD = 10.93$). In addition to appropriate reliance, students frequently followed misleading recommendations on the subset of misleading trials (overreliance: $M = 86.03\%$, $SD = 25.52$), while rejecting correct recommendations was uncommon (underreliance: $M = 2.37\%$, $SD = 8.27$). Post-task trust in the assistant was moderately high ($M = 5.01$, $SD = 1.23$; range: 1--7).

Trust was negatively associated with both appropriate reliance ($r = -.42$, $p < .001$) and task accuracy ($r = -.38$, $p < .001$). Correlations between appropriate reliance and AI literacy ($r = .03$, $p = .50$), need for cognition ($r = .04$, $p = .40$), programming self-efficacy ($r = .02$, $p = .65$), and programming literacy ($r = -.04$, $p = .46$) were not significant, suggesting that these characteristics may operate primarily by conditioning how trust relates to appropriate reliance, as examined in RQ2. Table~\ref{tab:corr} reports Pearson correlations among appropriate reliance, trust, and the individual difference measures. 

\begin{table}[t]
\centering
\caption{Pearson correlations (lower triangle). \textit{Note.} $^{*}p<.05$, $^{**}p<.01$, $^{***}p<.001$.}
\label{tab:corr}
\begin{tabular}{lcccccc}
\toprule
Variable & 1 & 2 & 3 & 4 & 5 & 6 \\
\midrule
1. Appropriate reliance (\%) &  &  &  &  &  &  \\
2. Post-task trust & -0.42$^{***}$ &  &  &  &  &  \\
3. AI literacy & 0.03 & 0.15$^{**}$ &  &  &  &  \\
4. Need for cognition & 0.04 & -0.00 & 0.32$^{***}$ &  &  &  \\
5. Programming self-efficacy & 0.02 & -0.04 & 0.38$^{***}$ & 0.51$^{***}$ &  &  \\
6. Programming literacy & -0.04 & 0.10$^{*}$ & 0.51$^{***}$ & 0.49$^{***}$ & 0.68$^{***}$ &  \\
\bottomrule
\end{tabular}
\end{table}

\subsection{RQ1: Trust and appropriate reliance during AI-assisted problem-solving}

\begin{figure}[t]
  \centering
  \begin{subfigure}[t]{0.49\linewidth}
    \centering
    \includegraphics[width=\linewidth]{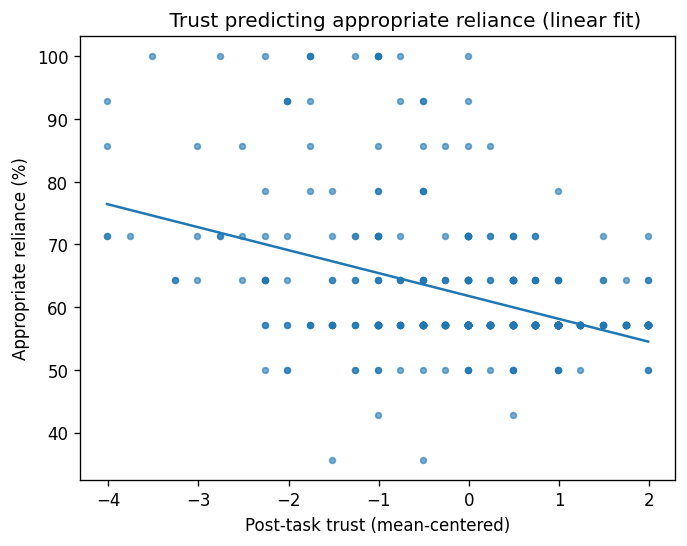}
    \caption{Linear fit.}
    \label{fig:rq1_linear}
  \end{subfigure}
  \hfill
  \begin{subfigure}[t]{0.49\linewidth}
    \centering
    \includegraphics[width=\linewidth]{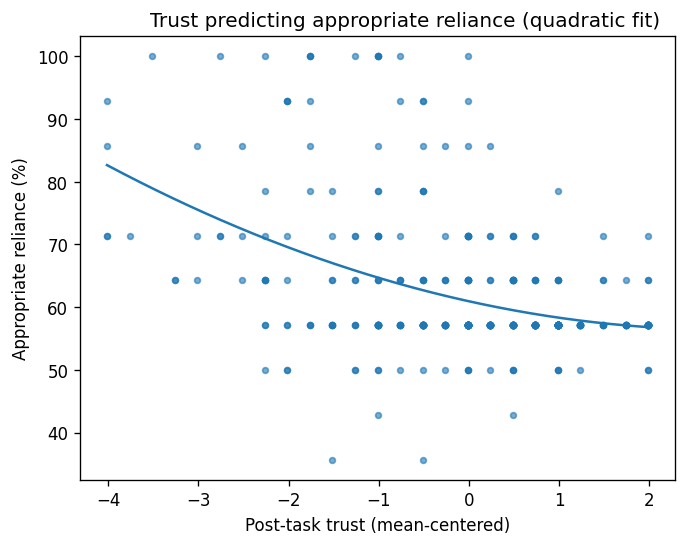}
    \caption{Quadratic fit.}
    \label{fig:rq2_quadratic}
  \end{subfigure}
  \caption{Post-task trust predicting appropriate reliance.}
  \label{fig:rq1_rq2_sidebyside}
\end{figure}

To estimate how trust related to students’ reliance on an AI assistant during problem-solving, we fit an ordinary least squares regression predicting appropriate reliance from post-task trust. Trust was a strong negative predictor of appropriate reliance ($b = -3.65$, $SE = 0.39$, $t(430) = -9.48$, $p < .001$, $R^2 = .173$), indicating that students who reported higher trust tended to make fewer decisions appropriately in this task. Given that the assistant intentionally included misleading recommendations, this negative association is consistent with the possibility that higher trust was linked to less scrutiny of recommendations before acceptance, resulting in less selective reliance overall. Figure~\ref{fig:rq1_linear} visualizes the linear fitted association.

\subsubsection{Testing for nonlinearity in the trust--reliance association.} In AI-assisted learning contexts, both underreliance (rejecting useful support) and overreliance (accepting flawed support) can undermine outcomes, but our task data suggest that underreliance was uncommon while overreliance was frequent. Prior trust and reliance research distinguishes these failure modes and emphasizes trust calibration, where learners appropriately rely on automated support when it is likely to be correct and withhold reliance when it is likely to incorrect \cite{pitts2025understandinghumanaitrusteducation,pitts2025students,zhang2020effect}. We therefore tested whether the trust--appropriate reliance association was nonlinear, consistent with the possibility that reliance varies across levels of trust (e.g. low, moderate, \& high).

For appropriate reliance, the quadratic model explained more variance than the linear model ($R^2 = .185$ vs.\ $R^2 = .173$; $\Delta R^2 = .012$). Both the linear trust term ($b = -3.18$, $SE = 0.43$, $t(429) = -7.45$, $p < .001$) and the squared term ($b = 0.56$, $SE = 0.22$, $t(429) = 2.55$, $p = .011$) were statistically significant. As shown in Figure~\ref{fig:rq2_quadratic}, the estimated decline in appropriate reliance was steeper from lower to moderate trust values and then attenuated at higher trust values; however, the fitted curve did not reverse direction within the observed data, providing limited evidence of a mid-range “optimal trust” pattern in this task context. Given that underreliance was uncommon and overreliance was prevalent, this pattern is consistent with the possibility that students’ trust in AI differs from standard trust in automation, with trust operating more as a general tendency to accept AI output than as a mechanism that supports calibration. This motivates RQ2, which examines whether learner characteristics help explain when trust is most strongly associated with appropriate reliance.

\subsection{RQ2: Individual differences moderating trust and appropriate reliance}

To understand whether learner characteristics moderate the influence trust has on appropriate reliance, we estimated moderation models predicting appropriate reliance from standardized trust ($Trust_z$), each standardized moderator ($Mod_z$), and their interaction ($Trust_z \times Mod_z$). Table~\ref{tab:mod} reports coefficients for AI literacy, need for cognition, programming self-efficacy, and programming literacy. Across models, trust remained negatively associated with appropriate reliance. Two interactions were statistically significant: AI literacy and need for cognition.

\begin{figure*}[t]
\centering
\begin{subfigure}[t]{0.49\linewidth}
    \centering
    \includegraphics[width=\linewidth]{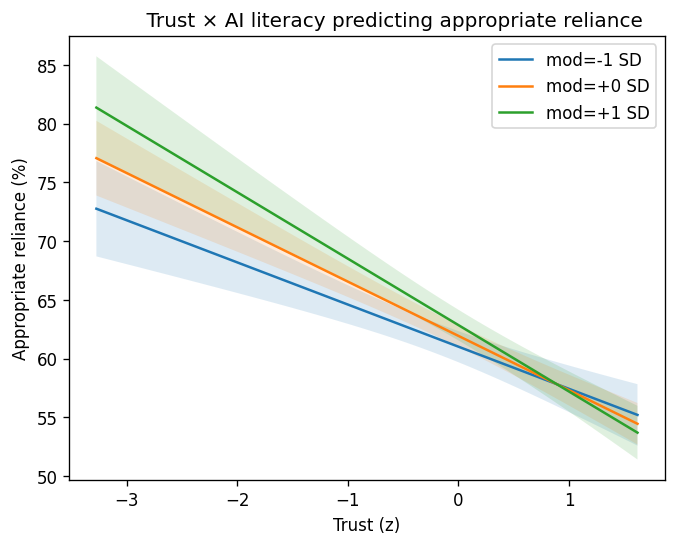}
    \caption{AI literacy}
    \label{fig:rq3_ai}
\end{subfigure}
\hfill
\begin{subfigure}[t]{0.49\linewidth}
    \centering
    \includegraphics[width=\linewidth]{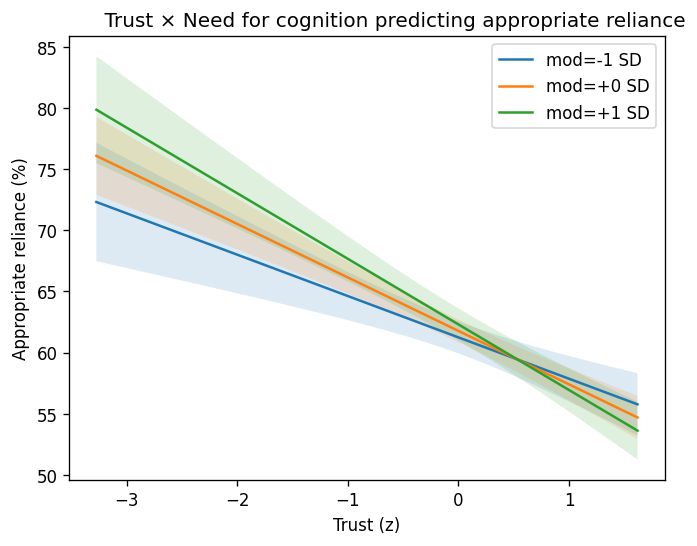}
    \caption{Need for cognition}
    \label{fig:rq3_nfc}
\end{subfigure}

\caption{Interaction models showing predicted appropriate reliance across trust at three levels of each moderator. Shaded bands represent 95\% confidence intervals.}
\label{fig:rq3_interactions}
\end{figure*}

\paragraph{AI literacy.}
The trust $\times$ AI literacy interaction was negative ($b = -1.03$, $p = .014$), meaning the estimated trust slope became more negative as AI literacy increased. Using the fitted coefficients, the conditional trust slope was approximately $-3.59$ at $-1$ SD AI literacy, $-4.62$ at the mean, and $-5.65$ at $+1$ SD. Figure~\ref{fig:rq3_ai} shows predicted values with 95\% confidence bands. At lower trust levels, higher AI literacy corresponded to higher predicted appropriate reliance, while this difference narrowed and slightly reversed at higher trust levels due to the steeper negative trust slope at higher AI literacy.

\begin{table}[t]
\centering
\caption{Moderation models predicting appropriate reliance from standardized trust, each moderator, and their interaction.}
\label{tab:mod}
\begin{tabular}{lrrrrrrrr}
\toprule
Moderator & $N$ & $R^2$ & $b_{Trust}$ & $p$ & $b_{Mod}$ & $p$ & $b_{Int}$ & $p$ \\
\midrule
AI literacy & 432 & .193 & -4.62 & $<.001$ & 0.92 & .054 & -1.03 & .014 \\
Need for cognition & 432 & .183 & -4.37 & $<.001$ & 0.53 & .263 & -0.99 & .040 \\
Programming self-efficacy & 432 & .178 & -4.33 & $<.001$ & 0.05 & .921 & -0.74 & .121 \\
Programming literacy & 432 & .174 & -4.43 & $<.001$ & 0.05 & .923 & -0.36 & .437 \\
\bottomrule
\end{tabular}
\label{table-2}
\end{table}

\paragraph{Need for cognition.}
Need for cognition also moderated the trust--appropriate reliance association ($b = -0.99$, $p = .040$). The negative interaction coefficient again implies that the estimated trust slope became more negative as need for cognition increased. Based on model estimates, the conditional trust slope was approximately $-3.38$ at $-1$ SD need for cognition, $-4.37$ at the mean, and $-5.36$ at $+1$ SD. Figure~\ref{fig:rq3_nfc} shows the predicted values with 95\% confidence bands. As with AI literacy, higher need for cognition corresponded to higher predicted appropriate reliance when trust was low, while the higher need for cognition line declined more steeply as trust increased.

\paragraph{Programming self-efficacy and programming literacy.}
The interactions for programming self-efficacy ($p = .121$) and programming literacy ($p = .437$) were not statistically distinguishable from zero, providing limited evidence in this task that these characteristics altered how trust related to appropriate reliance. As discussed in Section 5, this pattern may reflect the specific features of the present task and survey measures. Our study examined selective reliance during brief Python output-prediction problems, while the survey instruments captured broader perceptions of programming self-efficacy and literacy. Future work should continue examining these effects across other learning contexts.

\paragraph{Johnson--Neyman regions.}
To contextualize the significant interactions, we conducted Johnson--Neyman analyses to identify moderator values at which the conditional trust slope was statistically distinguishable from zero. For AI literacy, the Johnson--Neyman boundary occurred at $Mod_z \approx -2.37$, and for need for cognition at $Mod_z \approx -2.10$. Because these boundaries fall far into the lower tail of the observed moderator distributions, the conditional trust slope was statistically significant for most observed values of each moderator.

\section{Discussion}

This study examined how students’ trust in an AI assistant relates to their ability to rely on it selectively during programming problem-solving. Across analyses, higher trust was associated with lower appropriate reliance. Given that the assistant deliberately included misleading recommendations, this pattern suggests that students who trusted the AI more were less likely to distinguish between correct and incorrect suggestions and were more likely to accept the AI assistant's output without sufficient evaluation. This finding aligns with theoretical accounts of automation bias and cognitive offloading. When AI responses appear coherent and convincing, students may rely on quick judgments instead of carefully evaluating whether the recommendation is correct \cite{kahneman2011thinking}. In learning contexts, that shift can reduce metacognitive monitoring and self-explanation, making students less likely to scrutinize an AI-generated output before accepting it, especially when their trust in the AI system is high. Further, although the decline in appropriate reliance attenuated at higher levels of trust, the curve did not reverse within the observed range. This suggests that, in this context, moderate trust did not correspond to an “optimal” middle ground of calibrated reliance. Instead, trust operated primarily as a driver of overacceptance. This pattern differs from classic human–automation trust models, where both undertrust and overtrust are common \cite{lee2004trust}. Here, underreliance was rare and overreliance was frequent, indicating that learners’ interactions with generative AI is biased toward acceptance.

The moderation findings add nuance to the trust–reliance association found. The interaction plots showed that when trust was low, students higher in AI literacy and need for cognition had higher predicted appropriate reliance, suggesting that these characteristics may support more selective engagement with AI recommendations under lower-trust conditions. In both models, however, the interaction term was negative, indicating that the association between trust and appropriate reliance became more negative at higher levels of the moderator. As trust increased, the difference associated with higher AI literacy and need for cognition became smaller. These results suggest that these characteristics may help students evaluate AI suggestions more selectively when they approach the system with greater caution, but that this pattern weakens as trust rises. One possible explanation is that once students develop stronger trust in the system, even those who are generally more reflective or more familiar with AI may be less likely to verify suggestions in the moment.

For AI in education, these findings point to the importance of designing learning environments that make evaluation an explicit part of AI-supported work. Instructional activities and AI tools should make verification a routine part of task completion by prompting students to explain their reasoning or justify why they may agree or disagree with an AI-generated recommendation. These kinds of supports can help keep students engaged in the judgment process. Future work can build on this by developing systems that embed cognitive forcing functions into the interaction, such as requiring students to commit to an answer before viewing the AI recommendation, identify evidence for or against the suggestion, or complete a brief verification step before submitting a final response. Prior work has suggested that cognitive forcing functions can help reduce overreliance in AI-assisted decision-making, making them a useful direction for future educational systems that aim to support more careful and selective use of AI assistance \cite{de2025cognitive,vasconcelos2023explanations}. More broadly, these findings suggest that students' appropriate reliance on AI tools can be fostered through instruction and tool design. Future research can examine which forms of support, such as cognitive forcing functions, verification prompts, and staged response interfaces, are most effective for helping users question, check, and selectively use AI-generated recommendations during problem-solving.

\subsubsection{Limitations and Future Work.}

The study should be contextualized within its limitations. First, reliance patterns were bounded by our task structure. Participants completed 14 output-prediction items with a fixed mixture of correct and misleading recommendations, and the primary decision was whether to accept or reject the assistant’s guidance for each item. In longer assignments, reliance decisions can be distributed across many steps, including planning, debugging, and revision, where students may adopt AI output partially and iteratively. Future work should examine whether these reliance behaviors generalize to extended, open-ended tasks and should test instructional interventions that explicitly build verification routines and selective use of AI support. Second, our operationalization of appropriate reliance was tightly linked to task accuracy in this context. While this reflects successful discrimination between correct and incorrect AI suggestions, it does not directly capture students’ reasoning processes. Process-level data, such as think-aloud protocols or interaction logs, could help clarify whether students are actively evaluating or simply guessing correctly. Third, our measurement choices constrain interpretation. Trust and the individual difference factors were captured using self-report Likert scales aggregated at the participant level. While these measures showed strong internal consistency, they capture broad perceptions and tendencies and may not align perfectly with the specific skills this task required, such as verifying a recommendation under time pressure. In addition, our reliance measures were computed from observable accept versus reject decisions relative to recommendation correctness. This provides a clear behavioral operationalization, but it collapses different response strategies into a single score. For example, two students could achieve the same appropriate reliance by using different approaches, such as carefully checking each recommendation, relying on prior knowledge, or using superficial cues. Future work could pair these measures with additional indicators that clarify strategy use, such as brief justifications, confidence ratings, or response times.

\section{Conclusion}

This study examined how students’ trust in an AI assistant relates to their ability to rely on it selectively during programming problem-solving. Higher trust was consistently associated with lower appropriate reliance, indicating that students who trusted the system more were less likely to distinguish between accurate and inaccurate recommendations. Moderators of this relationship were identified, specifically, AI literacy and need for cognition moderated how trust related to appropriate reliance, suggesting that students’ understanding of AI and their tendency toward effortful thinking are relevant for how they engage with AI recommendations during problem-solving. These results motivate future work examining how appropriate reliance can be supported through instruction and tool design.

\section*{Data Availability Statement}

The anonymized data that support the findings of this study are available from the corresponding author, GP, upon reasonable request.

\bibliographystyle{splncs04}
\bibliography{ref}

@article{de2025cognitive,
  title={Cognitive Forcing for Better Decision-Making: Reducing Overreliance on AI Systems Through Partial Explanations},
  author={de Jong, Sander and Paananen, Ville and Tag, Benjamin and van Berkel, Niels},
  journal={Proceedings of the ACM on Human-Computer Interaction},
  volume={9},
  number={2},
  pages={1--30},
  year={2025},
  publisher={ACM New York, NY, USA}
}

@article{cheung2024reporting,
  title={Reporting reliability, convergent and discriminant validity with structural equation modeling: A review and best-practice recommendations},
  author={Cheung, Gordon W and Cooper-Thomas, Helena D and Lau, Rebecca S and Wang, Linda C},
  journal={Asia pacific journal of management},
  volume={41},
  number={2},
  pages={745--783},
  year={2024},
  publisher={Springer}
}

@inproceedings{rani2022investigating,
  title={Investigating the effects of different levels of user control on the effectiveness of context-aware recommender systems for web-based search},
  author={Rani, Neha and Chu, Sharon Lynn and Mei, Victoria Rene},
  booktitle={CHI Conference on Human Factors in Computing Systems Extended Abstracts},
  pages={1--6},
  year={2022}
}

@inproceedings{rani2025exploring,
  title={Exploring the Impact of User Feedback for Trust in Context-Aware Recommender Systems in Search-as-Learning},
  author={Rani, Neha and Kantamani, Srikar and Chu, Sharon Lynn},
  booktitle={International Conference on Human-Computer Interaction},
  pages={83--99},
  year={2025},
  organization={Springer}
}

@article{pitts2026drives,
  title={What Drives Students' Use of AI Chatbots? Technology Acceptance in Conversational AI},
  author={Pitts, Griffin and Motamedi, Sanaz},
  journal={arXiv preprint arXiv:2602.20547},
  year={2026}
}

@inproceedings{pitts2025survey,
  title={A Survey of LLM-Based Applications in Programming Education: Balancing Automation and Human Oversight},
  author={Pitts, Griffin and Hridi, Anurata Prabha and Narayanan, Arun Balajiee Lekshmi},
  booktitle={Proceedings of the Fourth Workshop on Bridging Human-Computer Interaction and Natural Language Processing (HCI+ NLP)},
  pages={255--262},
  year={2025}
}

@inproceedings{pitts2024proposed,
  title={A Proposed Model of Learners' Acceptance and Trust of Pedagogical Conversational AI},
  author={Pitts, Griffin and Marcus, Viktoria and Motamedi, Sanaz},
  booktitle={Proceedings of the Eleventh ACM Conference on Learning@ Scale},
  pages={427--432},
  year={2024}
}

@inproceedings{phung2025plan,
  title={Plan more, debug less: Applying metacognitive theory to AI-assisted programming education},
  author={Phung, Tung and Choi, Heeryung and Wu, Mengyan and Singla, Adish and Brooks, Christopher},
  booktitle={International Conference on Artificial Intelligence in Education},
  pages={3--17},
  year={2025},
  organization={Springer}
}

@inproceedings{ma2025generative,
  title={How Generative AI Impact Student Emotion and Engagement in Programming Tasks?},
  author={Ma, Boxuan and Guo, Liyuan and Yang, Tianyuan and Ding, Jihong},
  booktitle={International Conference on Artificial Intelligence in Education},
  pages={236--243},
  year={2025},
  organization={Springer}
}

@article{lee2004trust,
  title={Trust in automation: Designing for appropriate reliance},
  author={Lee, John D and See, Katrina A},
  journal={Human factors},
  volume={46},
  number={1},
  pages={50--80},
  year={2004},
  publisher={SAGE Publications Sage UK: London, England}
}

@inproceedings{zhang2020effect,
  title={Effect of confidence and explanation on accuracy and trust calibration in AI-assisted decision making},
  author={Zhang, Yunfeng and Liao, Q Vera and Bellamy, Rachel KE},
  booktitle={Proceedings of the 2020 conference on fairness, accountability, and transparency},
  pages={295--305},
  year={2020}
}

@article{tsai2019developing,
  title={Developing the computer programming self-efficacy scale for computer literacy education},
  author={Tsai, Meng-Jung and Wang, Ching-Yeh and Hsu, Po-Fen},
  journal={Journal of Educational Computing Research},
  volume={56},
  number={8},
  pages={1345--1360},
  year={2019},
  publisher={Sage Publications Sage CA: Los Angeles, CA}
}

@article{carolus2023mails,
  title={MAILS-Meta AI literacy scale: Development and testing of an AI literacy questionnaire based on well-founded competency models and psychological change-and meta-competencies},
  author={Carolus, Astrid and Koch, Martin J and Straka, Samantha and Latoschik, Marc Erich and Wienrich, Carolin},
  journal={Computers in Human Behavior: Artificial Humans},
  volume={1},
  number={2},
  pages={100014},
  year={2023},
  publisher={Elsevier}
}

@article{lins2020very,
  title={The very efficient assessment of need for cognition: Developing a six-item version},
  author={Lins de Holanda Coelho, Gabriel and HP Hanel, Paul and J. Wolf, Lukas},
  journal={Assessment},
  volume={27},
  number={8},
  pages={1870--1885},
  year={2020},
  publisher={Sage Publications Sage CA: Los Angeles, CA}
}

@inproceedings{jin2025learning,
  title={Learning by teaching: Enhancing music learning through llm-based teachable agents},
  author={Jin, Lingxi and Lin, Baicheng and Hong, Mengze and So, Hyo-Jeong and Zhang, Kun},
  booktitle={International Conference on Artificial Intelligence in Education},
  pages={148--155},
  year={2025},
  organization={Springer}
}

@article{choi2024utilizing,
  title={Utilizing generative AI for instructional design: Exploring strengths, weaknesses, opportunities, and threats},
  author={Choi, Gi Woong and Kim, Soo Hyeon and Lee, Daeyeoul and Moon, Jewoong},
  journal={TechTrends},
  volume={68},
  number={4},
  pages={832--844},
  year={2024},
  publisher={Springer}
}

@inproceedings{krupp2024challenges,
  title={Challenges and opportunities of moderating usage of large language models in education},
  author={Krupp, Lars and Steinert, Steffen and Kiefer-Emmanouilidis, Maximilian and Avila, Karina E and Lukowicz, Paul and Kuhn, Jochen and K{\"u}chemann, Stefan and Karolus, Jakob},
  booktitle={Proceedings of the International Conference on Mobile and Ubiquitous Multimedia},
  pages={249--254},
  year={2024}
}

@inproceedings{pitts2025students,
  title={Students’ reliance on ai in higher education: identifying contributing factors},
  author={Pitts, Griffin and Rani, Neha and Mildort, Weedguet and Cook, Eva-Marie},
  booktitle={International Conference on Human-Computer Interaction},
  pages={86--97},
  year={2025},
  organization={Springer}
}

@book{rani2023investigating,
  title={Investigating User Trust in Context-Aware Recommender Systems in Search as Learning},
  author={Rani, Neha},
  year={2023},
  publisher={University of Florida}
}

@inproceedings{pitts2025student,
  title={Student Perspectives on the Benefits and Risks of AI in Education},
  author={Pitts, Griffin and Marcus, Viktoria Medvedeva and Motamedi, Sanaz},
  booktitle={2025 ASEE Annual Conference \& Exposition},
  year={2025}
}

@inproceedings{lin2024factors,
  title={Factors Influencing College Students’ Use of AI Chatbots for Learning--Empirical study based on TAM extended model},
  author={Lin, Xiushui},
  booktitle={2024 5th International Conference on Artificial Intelligence and Electromechanical Automation (AIEA)},
  pages={151--159},
  year={2024},
  organization={IEEE}
}

@inproceedings{ehsan2021expanding,
  title={Expanding explainability: Towards social transparency in ai systems},
  author={Ehsan, Upol and Liao, Q Vera and Muller, Michael and Riedl, Mark O and Weisz, Justin D},
  booktitle={Proceedings of the 2021 CHI conference on human factors in computing systems},
  pages={1--19},
  year={2021}
}

@inproceedings{kizilcec2016much,
  title={How much information? Effects of transparency on trust in an algorithmic interface},
  author={Kizilcec, Ren{\'e} F},
  booktitle={Proceedings of the 2016 CHI conference on human factors in computing systems},
  pages={2390--2395},
  year={2016}
}

@article{von2021transparency,
  title={Transparency and the black box problem: Why we do not trust AI},
  author={Von Eschenbach, Warren J},
  journal={Philosophy \& Technology},
  volume={34},
  number={4},
  pages={1607--1622},
  year={2021},
  publisher={Springer}
}

@article{ranalli2021l2,
  title={L2 student engagement with automated feedback on writing: Potential for learning and issues of trust},
  author={Ranalli, Jim},
  journal={Journal of Second Language Writing},
  volume={52},
  pages={100816},
  year={2021},
  publisher={Elsevier}
}

@article{cacioppo1982need,
  title={The need for cognition.},
  author={Cacioppo, John T and Petty, Richard E},
  journal={Journal of personality and social psychology},
  volume={42},
  number={1},
  pages={116},
  year={1982},
  publisher={American Psychological Association}
}

@article{nazaretsky2025critical,
  title={The critical role of trust in adopting AI-powered educational technology for learning: An instrument for measuring student perceptions},
  author={Nazaretsky, Tanya and Mejia-Domenzain, Paola and Swamy, Vinitra and Frej, Jibril and K{\"a}ser, Tanja},
  journal={Computers and Education: Artificial Intelligence},
  pages={100368},
  year={2025},
  publisher={Elsevier}
}

@article{glikson2020human,
  title={Human trust in artificial intelligence: Review of empirical research},
  author={Glikson, Ella and Woolley, Anita Williams},
  journal={Academy of management annals},
  volume={14},
  number={2},
  pages={627--660},
  year={2020},
  publisher={Briarcliff Manor, NY}
}

@article{pitts2025understandinghumanaitrusteducation,
  title={Understanding human-AI trust in education},
  author={Pitts, Griffin and Motamedi, Sanaz},
  journal={Telematics and Informatics Reports},
  pages={100270},
  year={2025},
  publisher={Elsevier}
}

@article{parasuraman1997humans,
  title={Humans and automation: Use, misuse, disuse, abuse},
  author={Parasuraman, Raja and Riley, Victor},
  journal={Human factors},
  volume={39},
  number={2},
  pages={230--253},
  year={1997},
  publisher={SAGE Publications Sage CA: Los Angeles, CA}
}

@article{goddard2012automation,
  title={Automation bias: a systematic review of frequency, effect mediators, and mitigators},
  author={Goddard, Kate and Roudsari, Abdul and Wyatt, Jeremy C},
  journal={Journal of the American Medical Informatics Association},
  volume={19},
  number={1},
  pages={121--127},
  year={2012},
  publisher={BMJ Group BMA House, Tavistock Square, London, WC1H 9JR}
}

@article{sabharwal2023artificial,
  title={Artificial intelligence (ai)-powered virtual assistants and their effect on human productivity and laziness: Study on students of delhi-ncr (india) \& fujairah (uae)},
  author={Sabharwal, Dhruv and Kabha, Robin and Srivastava, Kajal},
  journal={Journal of Content, Community and Communication},
  volume={17},
  number={9},
  pages={162--174},
  year={2023}
}

@article{ahmad2023impact,
  title={Impact of artificial intelligence on human loss in decision making, laziness and safety in education},
  author={Ahmad, Sayed Fayaz and Han, Heesup and Alam, Muhammad Mansoor and Rehmat, Mohd and Irshad, Muhammad and Arra{\~n}o-Mu{\~n}oz, Marcelo and Ariza-Montes, Antonio and others},
  journal={Humanities and Social Sciences Communications},
  volume={10},
  number={1},
  pages={1--14},
  year={2023},
  publisher={Palgrave}
}

@book{kahneman2011thinking,
  title={Thinking, fast and slow},
  author={Kahneman, Daniel},
  year={2011},
  publisher={macmillan}
}

@article{vasconcelos2023explanations,
  title={Explanations can reduce overreliance on ai systems during decision-making},
  author={Vasconcelos, Helena and J{\"o}rke, Matthew and Grunde-McLaughlin, Madeleine and Gerstenberg, Tobias and Bernstein, Michael S and Krishna, Ranjay},
  journal={Proceedings of the ACM on Human-Computer Interaction},
  volume={7},
  number={CSCW1},
  pages={1--38},
  year={2023},
  publisher={ACM New York, NY, USA}
}

@article{li2023appropriate,
  title={Appropriate reliance on artificial intelligence in radiology education},
  author={Li, Matthew D and Little, Brent P},
  journal={Journal of the American College of Radiology},
  volume={20},
  number={11},
  pages={1126--1130},
  year={2023},
  publisher={Elsevier}
}

@article{duhaylungsod2023chatgpt,
  title={ChatGPT and other AI users: Innovative and creative utilitarian value and mindset shift},
  author={Duhaylungsod, Arvin V and Chavez, Jason V},
  journal={Journal of Namibian Studies},
  volume={33},
  number={2023},
  pages={4367--4378},
  year={2023}
}

@article{zhai2024effects,
  title={The effects of over-reliance on AI dialogue systems on students' cognitive abilities: a systematic review},
  author={Zhai, Chunpeng and Wibowo, Santoso and Li, Lily D},
  journal={Smart Learning Environments},
  volume={11},
  number={1},
  pages={28},
  year={2024},
  publisher={Springer}
}

@article{gaube2021ai,
  title={Do as AI say: susceptibility in deployment of clinical decision-aids},
  author={Gaube, Susanne and Suresh, Harini and Raue, Martina and Merritt, Alexander and Berkowitz, Seth J and Lermer, Eva and Coughlin, Joseph F and Guttag, John V and Colak, Errol and Ghassemi, Marzyeh},
  journal={NPJ digital medicine},
  volume={4},
  number={1},
  pages={31},
  year={2021},
  publisher={Nature Publishing Group UK London}
}

@article{buccinca2021trust,
  title={To trust or to think: cognitive forcing functions can reduce overreliance on AI in AI-assisted decision-making},
  author={Bu{\c{c}}inca, Zana and Malaya, Maja Barbara and Gajos, Krzysztof Z},
  journal={Proceedings of the ACM on Human-computer Interaction},
  volume={5},
  number={CSCW1},
  pages={1--21},
  year={2021},
  publisher={ACM New York, NY, USA}
}
\end{document}